# Student Perceptions of Large Language Models Use in Self-Reflection and Design Critique in Architecture Studio

Juan David Salazar Rodriguez[1], Sam Conrad Joyce[1], Nachamma Sockalingam[2], Khoo Eng Tat[3], Julfendi[4]

[1] META Design lab, Architecture and Sustainable Design Pillar, Singapore University of Technology and Design, Singapore, Singapore
[2] Office of Strategic Planning, Singapore University of Technology and Design, Singapore, Singapore
[3] College of Design and Engineering, National University of Singapore, Singapore, Singapore
[4] Engineering Product Development Pillar, Singapore University of Technology and Design, Singapore, Singapore

```
1008372@mymail.sutd.edu.sg, sam_joyce@sutd.edu.sg, na-
chamma@sutd.edu.sg, etkhoo@nus.edu.sg, julfendi_julf-
                endi@mymail.sutd.edu.sg
```

**Abstract.** This study investigates the integration of Large Language Models (LLMs) into the feedback mechanisms of the architectural design studio, shifting the focus from generative production to reflective pedagogy. Employing a mixed-methods approach with surveys and semi structured interviews with 22 architecture students at the Singapore University of Technology and Design, the research analyzes student perceptions across three distinct feedback domains: self-reflection, peer critique, and professor-led reviews. The findings reveal that students engage with LLMs not as authoritative instructors, but as collaborative "cognitive mirrors" that scaffold critical thinking. In self-directed learning, LLMs help structure thoughts and overcome the "blank page" problem, though they are limited by a lack of contextual nuance. In peer critiques, the technology serves as a neutral mediator, mitigating social anxiety and the "fear of offending". Furthermore, in high-stakes professor-led juries, students utilize LLMs primarily as post-critique synthesis engines to manage cognitive overload and translate abstract academic discourse into actionable design iterations.

## 1  Background and motivation

The architectural design studio is widely regarded as the discipline's signature pedagogy, a model historically rooted in the École des Beaux-Arts tradition and centered on learning through iterative production and public critique (Salama & El-Attar, 2010). Within this pedagogical paradigm, Donald Schön's (1984) concept of the reflective practitioner has been foundational, positioning design learning as a "reflective conversation with the situation." This reflective dialogue is typically enacted through three interconnected feedback channels: self-reflection, peer critique, and professor-led jury reviews. Together, these mechanisms are intended to support the development of critical judgment, design reasoning, and professional identity.

Despite their centrality, the effectiveness of these traditional feedback mechanisms has been increasingly questioned due to persistent psychosocial and cognitive barriers. The design jury, in particular, operates as a high-stakes, public performance that frequently induces anxiety rather than productive reflection. Instead of enabling reflection-in-action, students often experience juries as asymmetrical power encounters, prompting defensive postures and performative compliance rather than openness to critique (Webster, 2007). These dynamics undermine the pedagogical intent of the studio and contribute to broader well-being concerns within architectural education.

Peer critique, while theoretically positioned as a low-stakes and collaborative alternative, is similarly constrained by social friction. Hyland and Hyland (2019) describe a pervasive "fear of offending," whereby students prioritize social harmony over critical honesty. This phenomenon commonly referred to as the Mum Effect results in feedback that is overly positive, vague, or non-actionable, significantly diminishing its instructional value. Empirical studies further suggest that peer feedback is often distorted by friendship bias, where students hesitate to critique peers with whom they have close social ties or only value feedback from those perceived as artistically superior (McClean & Hourigan, 2013).

These socioemotional barriers are compounded by cognitive constraints inherent to architectural design. Design tasks are widely characterized as wicked problems, imposing a high intrinsic cognitive load due to their open-ended and ill-defined nature (Lavrsen & Daalhuizen, 2023). When students must simultaneously process large volumes of verbal feedback, manage contradictory critiques, and navigate social stressors, extraneous cognitive load increases substantially. Prior work has shown that such overload inhibits students' ability to synthesize, retain, and apply feedback, which turns into limiting deep learning (Mohamed et al, 2013).

In response to these long-standing challenges, architectural educators have experimented with a range of non-AI pedagogical interventions aimed at restructuring the social dynamics of critique. Frameworks such as the Critical Friends protocol redefine the role of the critic as a trusted interlocutor rather than an evaluator, formalizing feedback into "warm," "cool," and "hard" categories to reduce interpersonal friction (Costa & Kallick, 1993). Other strategies including anonymous peer review, silent critiques, gallery walks, and flipped classroom models seek to lower social risk, democratize participation, and shift critique from performative confrontation to reflective engagement (Baytiyeh, 2017; Francek, 2006; Miyazoe & Anderson, 2011). While these approaches demonstrate measurable benefits, they remain labor-intensive, context-dependent, and difficult to scale consistently across studio settings.

Concurrently, the rapid integration of Generative Artificial Intelligence (GenAI) and Large Language Models (LLMs) into architectural education has opened new avenues for pedagogical innovation. The dominant body of research has focused almost exclusively on LLMs as tools for generative production and technical optimization rather than as supports for pedagogical practices such as facilitation or reflective learning (Salazar Rodriguez et al., 2025). Studies have emphasized their role in early-stage ideation, rapid visual brainstorming, and overcoming creative inertia (Huh et al., 2025), as well as in convergent tasks such as structural code generation, compliance checking, and urban design automation (Naser, 2024).

Scholarship has only recently begun to examine LLMs as feedback-generating tools in educational contexts, but empirical understanding of their pedagogical role remains limited. While emerging work suggests that LLM-mediated feedback may reduce perceived power imbalances in settings such as doctoral supervision and academic writing (Tensen et al., 2025), there is little empirical insight into how LLMs might operate within the social and cognitive ecology of the design studio—not merely as generators of content, but as mediators of critique and scaffolds for reflection. Specifically, it remains unclear how students perceive LLMs as neutral interlocutors capable of bypassing the fear of offending in peer critique, or as cognitive aids capable of synthesizing fragmented and sometimes contradictory jury feedback.

Research in Human–Computer Interaction (HCI) shows that LLMs can mediate collaboration, conflict resolution, and feedback in ways that reduce social anxiety and defensiveness, but their potential to mitigate critique-related anxiety and cognitive overload in the architectural design studio remains underexplored. Studies on AI-mediated collaboration indicate that LLMs can function as psychological buffers, reducing social anxiety and encouraging participation by offering a non-judgmental interaction partner (Zhang et al., 2025). In conflict resolution and online moderation, LLMs have demonstrated a capacity to maintain perceived impartiality and de-escalate emotionally charged exchanges, often outperforming human moderators in perceived neutrality (Li et al., 2025). Additionally, work on machine-generated feedback suggests that receiving critique from an AI prior to human evaluation can lower ego defensiveness and increase receptivity to subsequent feedback (Zou et al., 2024). Despite these insights, their application to the architectural design studio particularly in mitigating critique-related anxiety and cognitive overload remains largely speculative.

At the same time, critical scholarship cautions against unreflective AI adoption. Empirical studies on creativity and LLM use indicate a risk of generative homogenization, where reliance on AI raises baseline performance but reduces collective diversity, leading to convergent design outcomes (Anderson et al., 2024). This tension underscores the need to conceptualize LLMs not as authoritative critics or replacements for human judgment, but as reflective partners whose role is carefully constrained and pedagogically aligned.

Against this backdrop, the present study investigates the role of LLMs as mediators within architectural design feedback processes. Rather than focusing on AI-driven production, the study centers on students' lived experiences of critique and reflection, and on how AI-mediated systems might restructure feedback interactions without displacing human agency. Specifically, the study addresses the following research questions:

1. How do architecture students currently engage with traditional feedback loops of self-reflection, peer critique, and professor-led reviews?
2. What are the perceived strengths and weaknesses of integrating LLMs into these distinct feedback domains?
3. What pedagogical roles and system features do students envision for LLMs to effectively support learning while preserving critical judgment?

This research responds to urgent concerns surrounding student well-being, feedback literacy, and cognitive overload in architectural education. It contributes a reframing of LLMs as partners in reflection rather than tools of production, proposing a hybrid feedback model in which AI supports low-level synthesis and emotional buffering while human educators retain responsibility for high-level conceptual and ethical mentoring. In doing so, the study advances a pedagogically grounded framework for the ethical and reflective integration of AI in the architectural design studio.

## 2    Methods

### 2.1    Research Design

This study employed a mixed-methods research design in which the quantitative findings informed subsequent qualitative analysis to investigate architecture students' reflections, perceived challenges, strengths, and expectations related to critique sessions when using Large Language Models (LLMs). The mixed-methods approach was selected to capture both the breadth of students' experiences across different critique contexts and the depth of their lived experiences within the architecture studio.

The quantitative phase consisted of three online surveys, each focusing on a distinct critique modality commonly encountered in architectural education: (1) self-reflection, (2) peer-to-peer critique, and (3) professor-led critiques. Across the three surveys, data were collected primarily through closed-ended questions, including Likert-scale items and multiple-choice questions, complemented by a limited number of open-ended questions that allowed participants to elaborate on specific responses. The surveys were administered online using Microsoft Forms and could be completed asynchronously at a time convenient for participants. All survey responses were securely stored within the principal investigator's institutional Microsoft account.

Following the survey phase, semi-structured interviews were conducted to further explore the same thematic domains, reflection practices, challenges, strengths, and expectations but in a more open and conversational manner. The interviews aimed to elicit richer accounts of students' experiences in the architecture studio and to gain insight into how future tools leveraging LLMs could better support critique processes. Interviews were audio-recorded with participants' consent and securely stored on the principal investigator's computer.

Consistent with a sequential explanatory design, survey data collection preceded the interviews, which were conducted several days later. Insights from the survey responses informed the focus of the interview discussions, allowing qualitative data to contextualize and expand upon patterns observed in the quantitative findings.

After data collection, quantitative survey data were analyzed using descriptive statistical techniques to identify trends across critique modalities, while qualitative data from both open-ended survey responses and interview transcripts were analyzed using an inductive thematic analysis approach. This combined analysis enabled the identification of recurring themes related to students' reflections, challenges, strengths, and expectations concerning critique practices and the use of LLMs.

The overarching goal of this mixed-methods design was to derive key design requirements and features for future tools whether standalone LLM-based systems or LLM-integrated platforms that could meaningfully enhance self-reflection, peer critique, and professor-led critique within the architectural design studio.

### 2.2 Participants and sampling

A total of 22 students enrolled at the Singapore University of technology and Design in the Architecture and Sustainable Design Pillar participated in the study, with an average age of 23.5 years. As illustrated in **Fig 1**, the gender distribution was nearly balanced, comprising 50.0% Female and 45.5% Male, with the remaining 4.5% preferring not to disclose their gender. Regarding the current level of architectural study, **Fig 2** indicates that the largest group of participants consisted of Year 3 students (45.5%), followed by Year 1 (22.7%) and Year 2 (13.6%), while Master's level students and those in other categories each accounted for 9.1%.

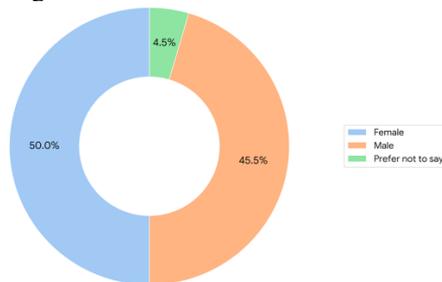
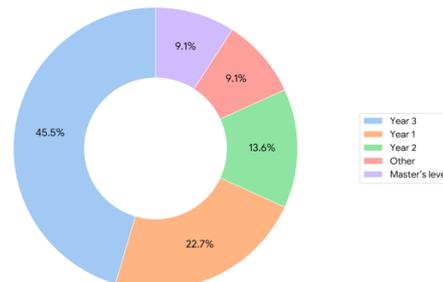

**Fig. 1. Distribution of Gender**    **Fig. 2. Current Level of Architectural Study**

Participation was subject to two inclusion criteria. First, participants were required to have prior experience with architectural studio critiques, including self-reflection, peer-to-peer critiques, or professor-led critiques. Second, participants were required to have used Large Language Models (LLMs) for any aspect of their architectural studio work, including reflection, critique preparation, or design-related tasks. These criteria ensured that respondents could meaningfully reflect on both critique practices and the role of LLMs within studio contexts.

Participants were recruited through institutional email announcements and university messaging application groups. Participation was entirely voluntary, and no academic or evaluative consequences were associated with participation or non-participation. Students who met the inclusion criteria and consented to participate were able to complete the surveys and, if selected, take part in the follow-up semi-structured interviews.

## 2.3 Data collection

The study employed a mixed-method data collection strategy designed to capture both the breadth of student experiences and the depth of their pedagogical engagement. The process was divided into two distinct phases: a quantitative survey series followed by qualitative semi-structured interviews.

**Survey Instruments.** Data collection commenced with the administration of three distinct surveys, designed to assess different pedagogical dimensions:
- Self-Learning (Appendix, Table 1)
- Peer- Learning (Appendix, Table 2)
- Professor-Led Learning (Appendix, Table 3)

Each survey instrument was structured around four core thematic sections: Reflections, Challenges, Strengths, and Expectations. The questions were predominantly closed-ended to establish a quantitative baseline of student sentiment, supplemented by selected open-ended questions to capture initial qualitative insights.

**Semi-Structured Interviews.** Following the completion of the surveys, participants were subjected to a semi-structured interview protocol (Appendix, Table 4). These interviews were designed to deepen the inquiry into the four themes established in the surveys.

## 2.4 Procedure

The study followed a sequential explanatory procedure, with quantitative data collection preceding qualitative data collection. Data collection began with the administration of three online surveys, each corresponding to a different critique modality: self-reflection, peer-to-peer critique, and professor-led critiques. The surveys were distributed simultaneously and administered via Microsoft Forms. Participants completed the surveys on their personal computers at a time of their convenience, with each survey requiring approximately 15 minutes to complete. Throughout the surveys, participants were encouraged to provide additional detail in open-ended questions to further elaborate on their experiences and perspectives.

Prior to accessing the surveys, participants were presented with an informed consent form detailing the objectives of the study, the voluntary nature of participation, potential risks, data handling procedures, and confidentiality measures. Only participants who provided consent were able to proceed with the surveys. Following the survey phase, a subset of participants was invited to participate in follow-up semi-structured interviews. Interview participants were selected from among survey respondents who met the inclusion criteria and expressed willingness to take part in an interview. The interviews were conducted several days after survey completion to support the sequential explanatory design.

Semi-structured interviews were conducted online via Zoom and lasted approximately 30–40 minutes on average. Participants were encouraged to expand upon their survey responses, provide concrete examples from their studio experiences, and elaborate on their reflections, challenges, strengths, and expectations regarding critique sessions and the use of Large Language Models. With participants' consent, interviews were audio recorded and subsequently transcribed for analysis. All recordings and transcripts were securely stored on the principal investigator's computer.

## 3 Results

### 3.1 Self-learning

**Reflections: Usage Patterns and The Role of AI**
The data suggests a distinct shift in how these tools are perceived not as authoritative instructors, but as collaborative aids. The vast majority of students, representing 58% of the sample group, perceive the LLM's primary role as a "Discussion partner," compared to only 16% viewing it as a "Tutor" and 6% as a "Critic"(See **Fig 3**). This aligns with interview findings where the respondent described using LLMs to check if ideas are clear and well-justified rather than seeking a final answer. Reflection is treated as an iterative process rather than a final checklist. Students utilize LLMs most frequently "during design iteration" (41%) and "after critiques" (37%), rather than solely before final submission (See **Fig 4**). In the context of high-stakes professor-led critiques, the reliance shifts toward synthesis. 95% of respondents identified the ideal role of AI as a "Post-critique reflection partner". This suggests students are using LLMs to deconstruct and process the dense feedback received during juries, rather than using AI to replace the professor's judgment.

**Qualitative Insight:** *"I see critique as an opportunity for me to improve... I often ask myself how the design responds to the design brief, as well as how clear and reasonable the concept is."*

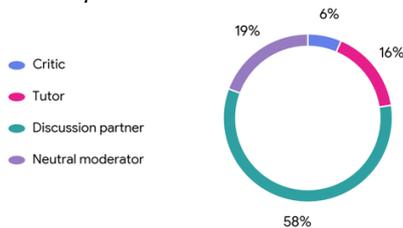 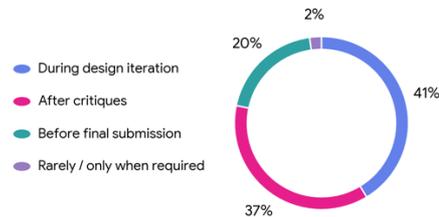

**Fig. 3. Perceived Role of the LLM**  **Fig. 4. Timing of LLM Usage in the Design Process**

**Challenges: The Specificity Gap and Human Uncertainty**
While adoption is high, students face significant challenges regarding the depth of AI feedback and their own confidence in evaluating architectural work. The most prevalent technical limitation is the lack of contextual nuance. 32% of respondents reported that LLM responses are "too generic," and 36% noted a "misunderstanding of design intent" (See **Fig 5**). The interview respondent reinforced this, stating that LLM feedback can be too general and might not fully capture the specifics of the project. The data reveals that the primary barrier to self-reflection is structural rather than emotional, with "Uncertainty about what to evaluate" (30%) being the most significant challenge. In contrast, "Emotional discomfort" is negligible at only 5%, suggesting students are willing to critique their work but simply lack the necessary framework to do so. This finding validates the potential utility of an LLM, which can directly address the other major hurdles by providing the specific vocabulary (20%) and objectivity (21%) that students are currently missing (See **Fig 6**).

**Qualitative Insight:** *"One challenge is that LLM feedback can be too general... There is also a risk of relying on it too much, so I always try to still largely use my own thinking and judgement."*

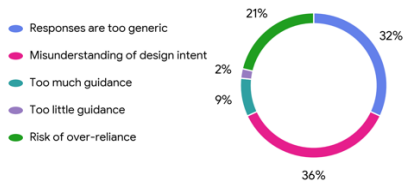 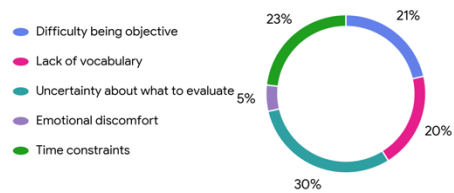

**Fig. 5. Reported Limitations of LLMs**   **Fig. 6. Top Barriers in Reflection**

**Strengths: Objectivity, Organization, and Social Safety**

The results indicated that the Large Language Models served primarily as a cognitive scaffolding tool rather than a replacement for deep design expertise. Participants overwhelmingly reported that the tool aided in the organization of their ideas, with 97% agreeing or strongly agreeing that it helped structure their thoughts. Furthermore, 92% of respondents noted that the LLM prompted them to consider new questions, effectively preventing "tunnel vision" during the reflection process (see **Fig 7**). These findings addressed the previously identified barrier of methodological uncertainty, suggesting that the LLM effectively mitigated the "blank page" problem by providing a necessary framework for self-evaluation. When compared directly to human feedback, the LLM demonstrated distinct advantages in communication and emotional safety, though it lagged in practical utility. A significant majority (85%) of participants rated the clarity of the LLM's feedback as superior to that of human tutors, reinforcing its role in supplying the descriptive vocabulary that students often lacked. Additionally, 65% of respondents preferred the emotional impact of the AI, likely due to its perceived objectivity, which 55% rated as better than human critique. However, the tool was viewed as less practically helpful; "usefulness" received the lowest positive rating at 45% (see **Fig 8**), with a notable portion of students finding it only comparable to or worse than human advice. This suggested that while the LLM was an effective, low-stakes partner for initiating reflection and reducing anxiety, it did not yet match the specific, high-level design guidance provided by human mentors.

**Qualitative Insight:** *"They are especially useful for giving general objective feedback and suggesting areas to focus on."*

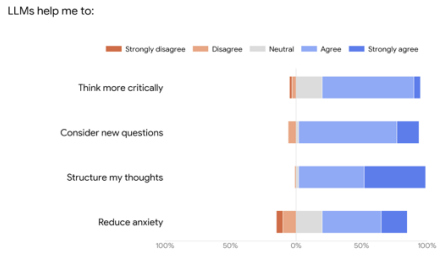
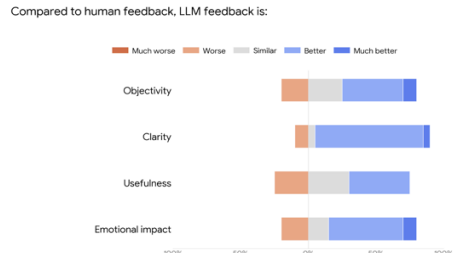

**Fig. 7.** LLMs perceived strengths in reflection

**Fig. 8.** LLM vs. Human Feedback

**Expectations: The Demand for Context-Aware Tools**

Participants expressed a strong preference for a collaborative approach to AI-mediated reflection. As illustrated in **Fig 9**, an overwhelming 86% of respondents indicated that the ideal LLM should offer "a balance of both" guidance and freedom, rejecting the extremes of total autonomy (9%) or rigid instruction (5%). This finding suggested that students valued the tool as a flexible partner capable of adapting to their workflow rather than a prescriptive tutor.

This demand for balanced support was operationalized in the specific feature requests detailed in **Fig 10**. The highest priorities were "Bias or objectivity checks" (15 students), followed by "Structured reflection frameworks" and "Studio-specific vocabulary" (13 students each). These preferences aligned with the previously identified barriers of subjectivity and lack of terminology, confirming that students sought an LLM that acted as a "cognitive mirror" one that provided structure and neutrality without imposing a specific design direction.

**Qualitative Insight:** *"Ideally, I see LLMs and other AI tools as support tools that help organize thinking and make feedback clearer. The main critique and design decisions should still come from people..."*

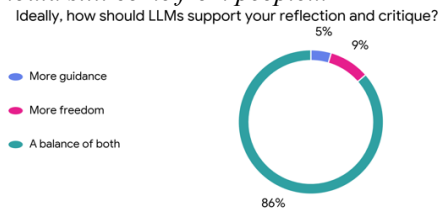
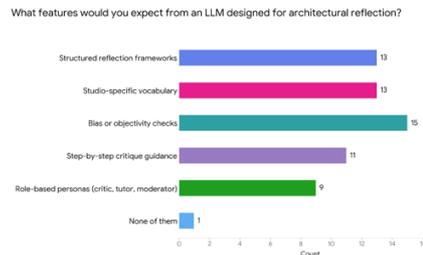

**Fig. 9.** Preferred Integration of LLMs in Reflection

**Fig. 10.** Most Desired LLMs Features

## 3.2   Peer learning

**Reflections: Utilization and the Social Context of Critique**

The survey results highlight a distinct contrast between the perceived value of peer feedback and the actual adoption of AI tools to support it. While 70% of respondents consider peer critiques "Important" or "Extremely Important" for their learning, the use of Large Language Models (LLMs) to *prepare* for these interactions remains surprisingly low. Only 10% of students reported using LLMs "Yes" (regularly) to prepare peer critiques, with another 15% using them "Sometimes." The vast majority (75%) do not use AI for preparation, which contrasts sharply with the ubiquitous use of AI for self-reflection found in other studio contexts (see **Fig 11**). Despite this low usage in preparation, the formats of critique suggest an environment ripe for intervention. Peer critiques are predominantly "Informal discussions" (85%) or "Pin-ups" (65%) (see **Fig 12**). In these unstructured settings, students perceive the LLM's potential role not as a replacement for peers, but as "Knowledge support" (60%) or a "Neutral mediator" (45%). This suggests that while students are not yet habitually using AI to prep for peer reviews, they recognize its potential to stabilize the often fluid and informal nature of student-to-student feedback.

**Qualitative Insight:** *"I also use them to better understand, organize, and respond to feedback from my peers... However, some of my peers view the use of LLMs and other AI tools negatively, so in such cases I will try to avoid the use of LLMs within that group project."*

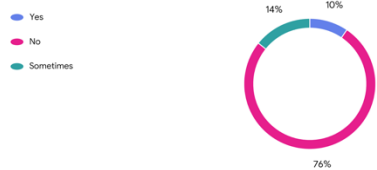
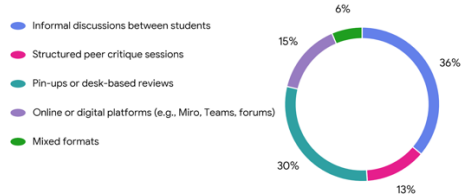

**Fig. 11. Frequency of LLM Usage for Peer Critique Preparation**  **Fig. 12. Common Formats of Peer Critique**

**Challenges: Social Anxiety and the "Shared Ignorance"**
The primary barriers in peer critique are deeply rooted in social dynamics and confidence gaps. A striking 27% of respondents identified "Lack of confidence in my expertise" as a major challenge when giving feedback, while 23% struggled with "Finding the right architectural vocabulary" (see **Fig 13**) This "shared ignorance," as one respondent phrased it, creates a feedback loop where students hesitate to critique because they feel unqualified or fear offending their peers (20%). On the receiving end, students struggle to filter the noise. 24% reported "Conflicting opinions" as a challenge, and 22% cited "Difficulty deciding what to act on." (see **Fig 14**) When LLMs are introduced to this equation, they bring their own set of challenges. The most significant limitation reported was "Over-generalized feedback" and a "Lack of contextual understanding". This indicates that while students struggle with the *social* delivery of feedback, AI currently struggles with the *contextual* specificity required to make that feedback relevant.

**Qualitative Insight:** *"Fear of giving feedback that students might not understand... [and] Shared ignorance."*

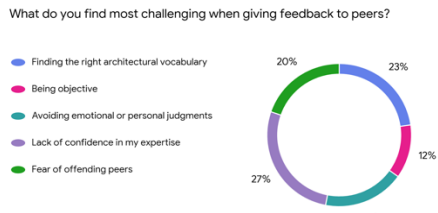
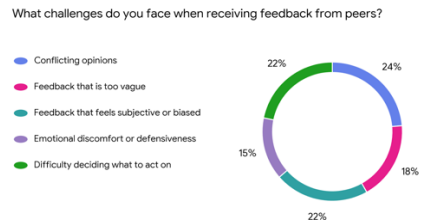

**Fig. 13. Human Barriers to Giving Peer Feedback**

**Fig. 14. Challenges in Receiving Feedback**

**Strengths: Articulation and Objective feedback focused on design**

Despite the low usage for preparation, students identify powerful potential benefits for AI in the peer review space, particularly in overcoming the vocabulary and confidence barriers mentioned above. An overwhelming 28% of respondents believe LLMs would be effective at "Helping articulate criticism clearly." This directly addresses students who struggle to find the right words. Furthermore, 23% noted that AI can "Offer multiple perspectives," and 20% believe it can "Support less confident students (see **Fig 15**)." **Fig 16** indicates that students perceive the LLM as a tool for democratization, capable of leveling the playing field in design studios. A clear majority agreed that its integration would result in "More equal participation" and "Reduced dominance by confident students." Most significantly, the data shows a strong consensus approximately 80% that the tool will foster a "Greater focus on design rather than personality" suggesting that the AI's objectivity effectively decouples the architectural evaluation from the student's social performance.

**Qualitative Insight:** *"Non bias opinions - clarity in different perspectives... There's no need to think about how to phrase the feedback."*

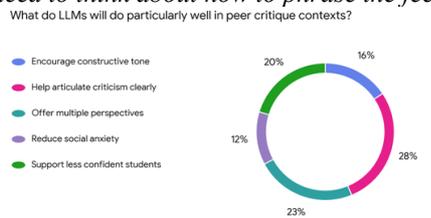
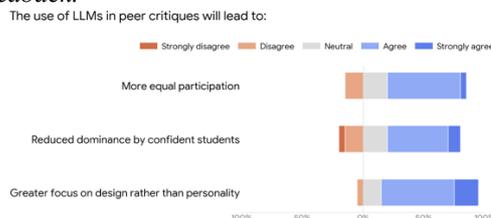

**Fig. 15. Perceived Strengths of LLMs in Peer Contexts**

**Fig. 16. Usage of LLMs in Peer Critique**

**Expectations: Scaffolding and Optional Integration**

Students envision a future where AI serves as a specialized scaffold rather than a mandatory moderator. There is a strong consensus on the need for domain-specific tools, with 75% of respondents requesting "Studio-specific vocabulary" and 65% asking for "Bias or subjectivity detection" (see **Fig 17**) The desire for bias detection highlights the students' awareness of the subjective nature of peer feedback and their desire for a tool that can parse helpful critique from personal preference. In terms of implementation, students prefer autonomy. 52% viewed the ideal integration of LLMs as an "Optional individual support tool," whereas only 22% supported the idea of a "Real-time moderator" during critiques (see **Fig 18**). This suggests that students want to use AI to bolster their own capabilities checking their bias, refining their vocabulary, and structuring their thoughts before and after engaging with their peers, rather than having an AI intervene in the conversation itself.

**Qualitative Insight:** *"Ideally, I see LLMs... as support tools that help organize thinking and make feedback clearer. The main critique... should still come from people."*

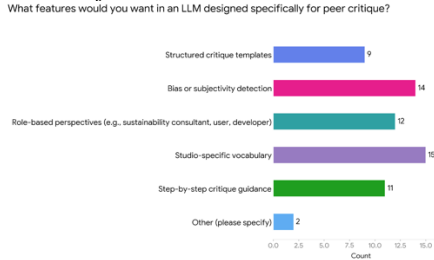

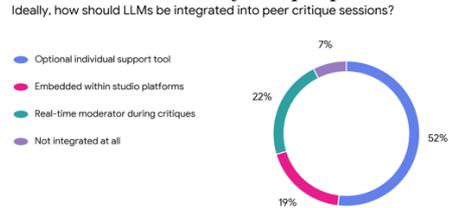

**Fig. 17.** Most Desired LLMs Features for Peer Critique

**Fig. 18.** Preferred Mode of LLM Integration in peer critique

### 3.3 Professor led learning

**Reflections: The Emergence of the AI Reflection Partner**

The survey indicates a significant integration of LLMs into the high-stakes environment of professor-led critiques, with 75% of students reporting that they use tools like ChatGPT to help interpret or reflect on feedback from professors. Unlike peer contexts where usage is lower, the pressure of the jury review drives students to seek support both before and after the event. 70% of respondents use LLMs to prepare for these sessions , primarily to "refine design narratives" (32%) (see **Fig 19**). However, the most dominant role for AI is retroactive; 95% of students identified the ideal role for LLMs as a "Post-critique reflection partner", helping them process the dense and often overwhelming information received during reviews. This usage pattern correlates with how students naturally manage critique data. 31% of students reflect by "reviewing comments later on" , and 25% revisit recordings or written notes (see **Fig 20**).

**Qualitative Insight:** *"Professors should still be the priority as they are the ones grading... [but AI] becomes a more democratic process, not just teacher saying X and student doing X."*.

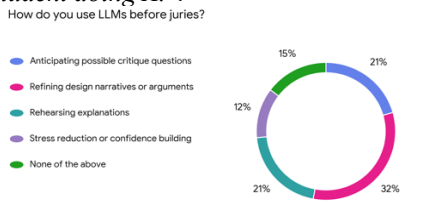

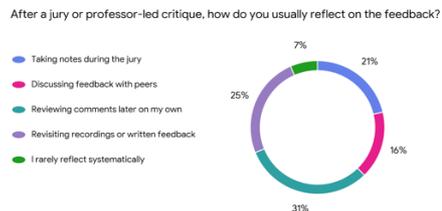

**Fig. 19.** LLM Usage in Jury Contexts

**Fig. 20.** Ideal Role of LLMs in Juries

**Challenges: Cognitive Overload and Interpretative Ambiguity**

The primary driver for AI adoption in this context appears to be the cognitive difficulty of processing real-time expert feedback. During the jury itself, 24% of students find it challenging to process "Conflicting opinions from different jurors", and 23% struggle with "Time pressure to respond". The complexity of the feedback often leads to confusion, with 23% of respondents noting difficulties with "References to unfamiliar precedents" (see **Fig 21**). These challenges persist long after the review ends. When trying to make sense of the feedback later, 23% of students struggle with "Ambiguous or abstract comments", and an equal 22% report a "Lack of clear next steps" (see **Fig 22**). The emotional and cognitive toll is significant, often leading to what one respondent described as a struggle to simply remember what was said while simultaneously trying to defend the work.

**Qualitative Insight:** *"I find it most challenging to explain confidently in front of [jurors]... [and] Understanding the bias of jurors."*.

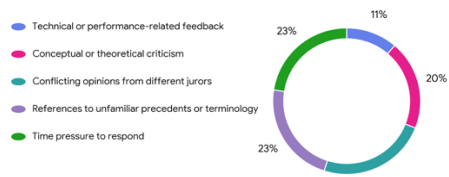
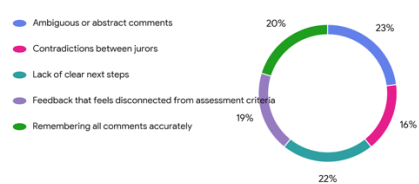

Fig. 21. Real-Time Challenges During Juries

Fig. 22. Post-Jury Processing Difficulties

**Strengths: Synthesizing Complexity and Bridging Gaps**

Students overwhelmingly value the LLM for its ability to impose order on chaos. In **Fig 23**, the highest-rated capability is "Help organize complex feedback". This expectation directly translates into practice: "Summarizing feedback" is the top reported use case in **Fig 24**. This suggests that the primary utility of the AI is not necessarily generating new ideas, but synthesizing the fragmented, unstructured data stream of a verbal jury into a coherent narrative. The data indicates a strong reliance on AI to decode expert language. "Bridge gaps in theoretical or technical knowledge" was the second-highest expected benefit in **Fig 23**. This is mirrored in **Fig 24**, where "Clarifying terminology or references" and "Translating abstract critique into concrete actions" are dominant activities. Students are essentially using the LLM as an on-demand tutor to demystify high-level architectural discourse and convert abstract academic feedback into actionable design steps.

**Qualitative Insight:** *"It would be great if the AI tools are able to record the feedback... Making feedback more organized - provide possibilities."*.

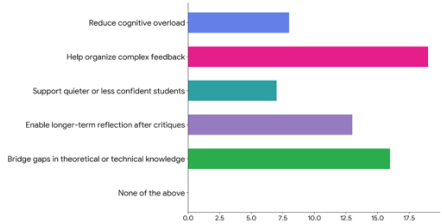 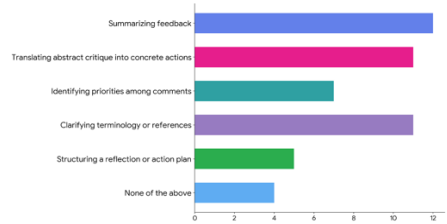

**Fig. 23. Perceived Strengths of LLMs in Jury Learning**    **Fig. 24. Specific Post-Jury LLM Tasks**

**Expectations: Structure, Alignment, and Actionability**

According to **Fig 25**, half of the participants (50%) view the ideal role of an LLM as a "Post-critique reflection partner," significantly outweighing its use as a preparatory tool (30%) or a real-time moderator (20%). Notably, zero participants selected "No role," indicating universal acceptance of AI integration in some form. This desire for retrospective support is driven by a need to organize chaotic feedback. **Fig 26** shows that the most requested features are a "Structured critique breakdown" (17 counts) and "Action-oriented summaries" and "Alignment with learning objectives or grading criteria" (14 counts). This suggests students value the AI's ability to translate unstructured verbal commentary into clear, actionable steps for design iteration.

**Qualitative Insight:** "*Help break down the feedback into less philosophical points... Give a personal summary of what the juror is trying to project*".

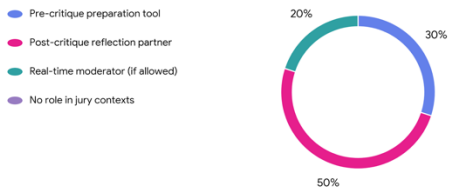 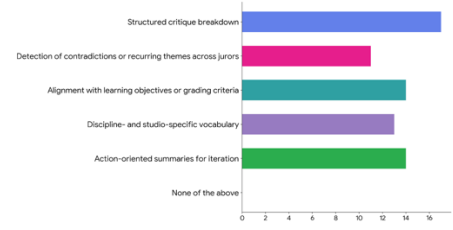

**Fig. 25. LLM preferred role in Professor-led critique**    **Fig. 26. Desired Features for Jury Support LLMs**

### 3.4  Semistructured interviews

Participants describe their engagement with Large Language Models (LLMs) not as a passive reception of information, but as an interactive, iterative dialogue where the tool serves primarily to clarify thoughts and structure complex ideas. One student explained, "I usually treat it as a chat partner to bounce ideas off. It helps me to articulate what I am thinking". This interaction is often compared to debugging in other fields, with another participant noting, "It is like a rubber duck debugging but for design. I explain my concept to it to see if it makes sense". The process is deeply reflective, as students use the tool "to check if my ideas are clear and well-justified. Sometimes I write down my messy thoughts and ask it to summarize the key points". To get the most out of these interactions, students often employ persona-based prompting: "I usually provide a persona, like 'act as a strict architectural critic' or 'act as a user of this building' to get different perspectives".

Despite the utility of these interactions, the most significant barrier reported by students is the LLM's inability to process visual and spatial information, leading to feedback that feels disconnected from the physical reality of the project. As one respondent described, "The main challenge is that it cannot 'see' my design. I have to describe everything in text, which is very hard for spatial concepts". This limitation often results in a lack of depth, with students noting that "Sometimes LLMs give very generic feedback that could apply to any project. It lacks the nuance of a real studio critique". Furthermore, there is a persistent concern about accuracy and dependency: "It hallucinates sometimes, giving me references to buildings that don't exist or misinterpreting architectural theories". Consequently, students feel they must remain vigilant, stating, "There is a risk of relying on it too much... I have to remind myself that it doesn't actually 'know' architecture".

In spite of these limitations, students highly value LLMs for their emotional neutrality and their ability to organize chaotic information, viewing the tool as a safe space for critique and cognitive offloading. One participant highlighted the benefit of this neutrality: "It is more objective... it doesn't have the bias that peers might have. Peers sometimes just say 'it's nice' to be polite". This objectivity helps mitigate the anxiety often associated with academic feedback: "I feel less anxious asking 'stupid' questions to an AI than to my professor". Beyond emotional support, the speed and organizational capacity of the tool are key strengths. One student remarked, "It helps organize thinking and make feedback clearer. Especially after a jury, I paste the transcript and ask it to find the main issues". Another emphasized the convenience, noting, "It is very fast. I can get feedback instantly at 3 am when I am working, which is impossible with professors".

Looking forward, students articulate a clear desire for tools that are integrated into their visual workflows and possess specialized architectural knowledge to bridge the current gaps. The most requested feature is visual understanding: "I hope it can understand images so I can just show my render or plan and get feedback on the spatial quality". Students also envision a seamless integration into their design software: "Ideally, it would be a plugin for Rhino or Revit, so it can see the model as I build it". Finally, there is a demand for more domain-specific accuracy, with students requesting features such as a "case study search that actually finds relevant architectural precedents based on my description" and a better grasp of terminology, as "It needs to know architectural jargon better. Sometimes it uses words in a non-architectural way".

## 4    Discussion

### 4.1    Cognitive skills

The following section synthesizes students' experiences with LLMs across different cognitive levels, structured through Bloom's Taxonomy, to clarify where LLMs meaningfully supports architectural learning and where its limitations remain evident.

At the foundational level of remembering, LLMs acts primarily as a post-event archival tool that mitigates the cognitive overload of critique sessions. Students frequently reported an inability to retain the sheer volume of information presented during reviews, with one respondent noting the difficulty of "remembering everything" and "catching references I never heard of". In this capacity, AI helps by serving as an external memory bank, converting ephemeral verbal feedback into permanent, reviewable records. Students envision future tools that can "record the feedback clearly" or provide "meeting minute summaries", effectively offloading the burden of memorization so they can focus on the discussion itself. However, LLMs currently fails to assist in the immediate moment of remembering due to processing latency. It does not yet help students "process other ongoing conversations and information in real-time". Until AI tools can listen and transcribe instantly without manual intervention, they remain retrospective archives rather than real-time cognitive aids, leaving students to manage the immediate stress of the critique unaided.

Moving to the level of understanding, LLMs serve as a powerful bridge between the novice student and the expert juror. Architectural education is often laden with complex terminology that can alienate students; respondents described jurors using "flowery language" or obscure references. LLMs helps significantly here by "explaining terms used" and "bridging the gap between technical expertise and beginner knowledge" This democratizes access to high-level theory, allowing students to grasp concepts that would otherwise be lost. Yet, this understanding is often superficial because LLMs lacks deep, studio-specific context. It operates as a dictionary rather than a cultural interpreter, often failing to grasp the "essence of studio concepts unless the prompt is really specific". Consequently, while LLM can define a term, it often misses the specific pedagogical intent or the "why" behind a professor's comment, leading to a disconnect between the definition and its application in the studio.

In the application phase, students successfully employ LLMs to structure their workflows and articulate their design narratives. The data indicates that LLMs are rarely used to generate the design itself but is invaluable for "structuring thoughts and discussions into a safe base or skeleton outline". It aids students in "better articulating and translating thoughts", effectively allowing them to rehearse their arguments before high-stakes presentations. This scaffolding gives students a "safe base" to test their explanations. However, LLMs hinder application when it is relied upon too heavily, often yielding "generic responses". Students found that while LLMs can organize an argument, it cannot apply design changes to the project itself; it can talk about architecture but cannot "do" it. Over-reliance here leads to a phenomenon where the verbal presentation improves while the actual design work remains static or disconnected from the narrative.

At the level of analysis, LLMs provide a unique benefit by stripping away social and emotional noise. Students value it for "filtering out harsh critiques" and offering "pure objectivity". This allows students to analyze feedback dispassionately, separating their personal feelings from the critique of their work, which is particularly difficult in the heat of a public review. Conversely, LLMs fail to provide the deep, intuitive analysis that comes from human experience. It cannot replace "genuine human feedback with years of unique human experience" because it cannot visually or spatially analyze the design. This limitation leads to a "shared ignorance" where the LLMs might hallucinate a critique that does not align with the visual evidence, proving that while it is an excellent text analyst, it remains a poor design critic.

When evaluating their own work, students use LLMs to create a psychological safety net. It serves as a non-judgmental "side buddy" that validates ideas and reduces the anxiety that often paralyzes decision-making. This "space and freedom to have conversations without being judged" is crucial for building confidence. Despite this, LLMs are not accepted as a valid final evaluator of architectural quality. Students explicitly state that "feedback from professors is still more important" and do not trust LLMs to make the final call on merit. While it helps students evaluate their options, it is viewed as inferior to the "human element" of judgment, which is seen as indispensable for assessing the true value of a design.

Finally, at the level of creation, LLMs help by acting as a boundless library that acts as a catalyst for ideation. Students envision it aiding creativity by "identifying real-life project references" and suggesting "branches of ideas generation", which is particularly useful when they are stuck. However, this comes with a significant caveat: the risk of standardization. There is a palpable fear among students that using LLMs will lead to "repetitive" outcomes and "less creativity" because the feedback is based on average data patterns rather than unique, individualistic thought. While AI helps generate a higher volume of ideas, it does not necessarily support the creation of better or more original ones, potentially leading to a homogenization of student work.

### 4.2 Key design features for future LLMs

Regarding *Self-Reflection*, students identified a critical disconnect where they design visually but are forced to reflect verbally. Future LLMs must bridge this gap through Multimodal Fluency. To address this, students requested image-to-text analysis tools capable of visual hierarchy analysis, essentially allowing them to input images and receive image-based outputs. A key feature would be an AI that acts as an objective "mirror," scanning a student's drawing to answer whether the visual rhetoric actually conveys their stated intent, enabling them to verify their internal narrative. Furthermore, instead of simply providing answers, the AI should employ Socratic prompting engines. This "Socratic mode" would challenge the student's logic with "what if" scenarios and simulate a desk critique to deepen critical thinking without offering direct solutions.

In the context of *Peer Critique*, the process is often hindered by social anxiety and a lack of vocabulary, meaning future tools should focus on Social Scaffolding and Bias Mitigation. Students expressed a strong desire for real-time bias detection, effectively a "tone check" that flags overly aggressive or non-constructive language before a peer sends it. Conversely, this tool could help shy students articulate criticism clearly to ensure their valid points are heard. To further overcome the fear of offending peers, collaborative anonymity was suggested. This feature would allow peers to input feedback anonymously, which the AI would then synthesize into a single "studio consensus" report, stripping away personality politics to focus purely on emerging design patterns.

Finally, for *Professor-Led Critique*, which acts as a high-information and high-stress event, the primary design needs are Cognitive Offloading and Contextual Linking. Students explicitly requested live transcription with speaker diarization—an AI that can listen in real time, recognize who is speaking, and separate a professor's feedback from a peer's comment. This would automatically generate "meeting minutes," freeing students from note-taking so they can maintain engagement and eye contact. Additionally, to address the pain point of professors citing unknown architects, an active referencing system is needed. This "Active Linker" would listen for proper nouns and immediately push visual references to the student's device, instantly filling knowledge gaps. Unlike current chat models that reset sessions, students also want longitudinal progress tracking with "long-term memory." This system would remember feedback from prior reviews and check if the current design has addressed those specific points, acting as a continuity guard against forgotten advice.

## 5      Conclusions

The integration of Large Language Models into the architectural design studio represents a fundamental shift from generative assistance to cognitive scaffolding. Through the lens of Bloom's Taxonomy, the data reveals that students are not primarily using AI to replace the creative act of design, but rather to navigate the complex, high-stakes social environment of the critique. AI currently functions most effectively at the foundational levels of Remembering and Understanding where it acts as a digital scribe and a jargon decoder, democratizing access to architectural theory and reducing the cognitive load of high-pressure reviews.

However, a significant gap remains between the text-based capabilities of current LLMs and the visual, multimodal nature of architectural education. While AI provides a psychological safety net that fosters confidence and objectivity in self-reflection, it currently fails to offer the deep, contextual, and spatial nuance required for high-level evaluation. Students view AI as a valuable "side buddy" for articulation and organization but explicitly reject it as a substitute for the "human element" of judgment provided by peers and professors.

Moving forward, the development of AI tools for architecture must transcend generic text interfaces to become context-aware studio partners. The feedback from this study calls for a new class of multimodal agents capable of real-time listening, visual-verbal alignment, and longitudinal progress tracking. By designing tools that act as objective mediators rather than passive generators, AI can evolve from a simple productivity tool into a sophisticated pedagogical instrument that supports the unique iterative process of the design studio.

# APPENDIX

Table 1. Self-reflection in Architecture studio with LLMs survey questions.

| Question | Category | Question Type |
|---|---|---|
| When do you usually reflect on your design work? (Select all that apply) | Reflections | Multiple Choice |
| What kinds of questions do you usually ask yourself when reflecting on your design? | Reflections | Open Ended |
| Which types of critique do you regularly engage in? | Reflections | Multiple Choice |
| Do you use LLMs to support self-reflection on your design work? | Reflections | Single Choice |
| How do you use LLMs during self-reflection? (Select all that apply) | Reflections | Multiple Choice |
| How would you describe your typical prompts? | Reflections | Single Choice |
| What do your reflection prompts usually focus on? | Reflections | Multiple Choice |
| Do you use LLMs to simulate or support design critiques? | Reflections | Single Choice |
| How do you mostly perceive the LLM's role? | Reflections | Multiple Choice |
| Has using LLMs changed how you learn or think about your design work? | Reflections | Single Choice |
| Overall, how do you feel about using an AI tool to reflect on your own thinking and design decisions? | Reflections | Open Ended |
| How difficult do you find critique sessions overall? | Challenges | Close Ended (Likert Scale) |
| What do you find most challenging about critique sessions? | Challenges | Open Ended |
| What challenges do you face when reflecting on or critiquing your own work? | Challenges | Multiple Choice |
| What difficulties have you experienced when using LLMs? | Challenges | Multiple Choice |
| Which other challenges do you face when using LLMs or other AI tools for self-reflection? | Challenges | Open Ended |
| LLMs help me to: | Strengths | Close Ended (Likert Scale) |
| Which are the strengths you perceive of using LLMs or AI tools for self-reflection? | Strengths | Open Ended |
| Compared to human feedback, LLM feedback is: | Strengths | Close Ended (Likert Scale) |
| Ideally, how should LLMs support your reflection and critique? | Expectations | Single Choice |
| What features would you expect from an LLM designed for architectural reflection? | Expectations | Multiple Choice |
| Please specify more features you can think of | Expectations | Open Ended |

**Table 2.** Peer critique in Architecture studio with LLMs survey questions.

| Question | Category | Question Type |
|---|---|---|
| How do peer critiques usually take place in your design studio? (Select all that apply) | Reflections | Multiple Choice |
| Please briefly describe how peer critiques typically work in your studio. (Open-ended) | Reflections | Open Ended |
| Compared to feedback from instructors, how important are peer critiques for your learning? | Reflections | Close Ended (Likert Scale) |
| Do you use Large Language Models (e.g., ChatGPT) to help prepare peer critiques? | Reflections | Single Choice |
| How do you use LLMs when preparing peer critiques? (Select all that apply) | Reflections | Multiple Choice |
| Have you used LLMs during or immediately after peer critique sessions? | Reflections | Single Choice |
| In peer critique situations, how do you mostly perceive the LLM's role? | Reflections | Multiple Choice |
| When prompting LLMs to critique a peer's design, what do you usually focus on? (Select all that apply) | Reflections | Multiple Choice |
| Do you modify LLM-generated feedback before sharing it with peers? | Reflections | Single Choice |
| How do you feel about learning from peers when AI tools are part of the critique process? | Reflections | Open Ended |
| What do you find most challenging when giving feedback to peers? Select all that apply | Challenges | Multiple Choice |
| What challenges do you face when receiving feedback from peers? Select all that apply | Challenges | Multiple Choice |
| Which other challenges do you face when you have peer critiques among students | Challenges | Open Ended |
| Where do LLMs fall short? select all that apply | Challenges | Multiple Choice |
| What kinds of insights do you usually gain from peer critiques? (Select all that apply) | Strengths | Multiple Choice |
| Using LLMs in peer critiques will make feedback: | Strengths | Close Ended (Likert Scale) |
| The use of LLMs in peer critiques will led to: | Strengths | Close Ended (Likert Scale) |
| What do LLMs will do particularly well in peer critique contexts? select all that apply | Strengths | Multiple Choice |
| What are the perceived strengths of using LLMs or other AI tools for peer critiques? | Strengths | Open Ended |
| Ideally, how should LLMs be integrated into peer critique sessions? | Expectations | Single Choice |
| What features would you want in an LLM designed specifically for peer critique? | Expectations | Multiple Choice |
| Which other features would you like LLMs or AI systems to integrate for having a peer critique? | Expectations | Open Ended |

**Table 3.** Professor-led critique in Architecture studio with LLMs survey questions.

| Question | Category | Question Type |
|---|---|---|
| How would you describe the typical format of juries or professor-led critiques you have experienced? (Select all that apply) | Reflections | Multiple Choice |
| In your experience, professor-led critiques and juries are primarily for: | Reflections | Close Ended (Likert Scale) |
| Juries typically make me feel: | Reflections | Close Ended (Likert Scale) |
| After a jury or professor-led critique, how do you usually reflect on the feedback? (Select all that apply) | Reflections | Multiple Choice |
| Do you use Large Language Models (e.g., ChatGPT) to help interpret or reflect on feedback from professors or jurors? | Reflections | Single Choice |
| How do you use LLMs after juries? (Select all that apply) | Reflections | Multiple Choice |
| Do you use LLMs to prepare for professor-led critiques or juries? | Reflections | Single Choice |
| How do you use LLMs before juries? (Select all that apply) | Reflections | Multiple Choice |
| Has using LLMs changed how you learn from juries or professor feedback? | Reflections | Single Choice |
| How do you feel about learning from professors when AI tools are part of the critique and reflection process? | Reflections | Open Ended |
| During juries, what do you find most difficult to process in real time? (Select all that apply) | Challenges | Multiple Choice |
| What do you find most challenging when making sense of jury feedback afterward? | Challenges | Multiple Choice |
| I have concerns about using LLMs in high-stakes critique settings | Challenges | Close Ended (Likert Scale) |
| What do you find most challenging of critique sessions? | Challenges | Open Ended |
| What will do LLMs do particularly well when supporting learning from juries? select all that apply | Strengths | Multiple Choice |
| What will be the strengths of using LLMs or AI tools during a critique session? | Strengths | Open Ended |
| Where should the boundary be between professor judgment and AI support in juries? | Expectations | Single Choice |
| Ideally, what role should LLMs play in professor-led critiques or juries? | Expectations | Single Choice |
| What features would you want in an LLM designed specifically to support learning from juries? | Expectations | Multiple Choice |
| Please specify more features besides the above mentioned | Expectations | Open Ended |

**Table 4.** Semi-structured interview LLMs in architecture studio critique questions.

| Question | Category |
|---|---|
| Can you briefly describe your current stage of architectural study and the types of design studios or projects you are working on? | Reflections |
| When working on a design project, how do you usually reflect on your work and engage with critique (Give examples)? | Reflections |
| Have you used any Large Language Models (e.g., ChatGPT) in your design studies? | Reflections |
| If you use LLMs, how do you typically use them to reflect on your own design work (Give examples)? | Reflections |
| Do you use LLMs to support design critiques—your own, peer critiques, or professor-led feedback? | Reflections |
| How do you usually prompt LLMs when using them for reflection or critique (Give examples)? | Reflections |
| How does feedback from an LLM compare with feedback from peers or professors? | Strengths |
| What challenges or concerns have you experienced when using LLMs for reflection or critique? | Challenges |
| What do you think LLMs do particularly well in supporting your learning for the Design studio? | Strengths |
| Do you think the use of LLMs can affect how students interact during critiques with peers or professors? | Strengths |
| Ideally, what role would you like LLMs to play in architectural reflection and critique? | Expectations |
| If an LLM were designed specifically for architecture studios critique, what features would you want? | Expectations |
| Through which platforms or channels can you imagine these interactions with LLMs for critiques (Miro board, webpages, messaging apps, LLMs interfaces, etc)? | Expectations |